\documentclass{PoS}

\title{Chiral symmetry restoration and strong CP violation in a strong magnetic background}

\ShortTitle{Chiral symmetry restoration and strong CP violation in a strong magnetic background}

\author{\speaker{Eduardo S. Fraga}
\\
        Instituto de F\'\i sica, Universidade Federal do Rio de Janeiro,\\
Caixa Postal 68528, 21941-972, Rio de Janeiro, RJ , Brazil \\
        E-mail: \email{fraga@if.ufrj.br}}

\author{Ana J\' ulia Mizher\\
        Instituto de F\'\i sica, Universidade Federal do Rio de Janeiro,\\
Caixa Postal 68528, 21941-972, Rio de Janeiro, RJ , Brazil \\
        E-mail: \email{anajulia@if.ufrj.br}}

\abstract{Motivated by the phenomenological scenario of the chiral magnetic effect that can be 
possibly found in high-energy heavy ion collisions, we study the role of very intense magnetic fields 
and strong CP violation in the phase structure of strong interactions and, more specifically, their 
influence on the nature of the chiral transition. Direct implications for the dynamics of phase conversion 
and its time scales are briefly discussed. Our results can also be relevant in the case of the early universe.}

\FullConference{5th International Workshop on Critical Point and Onset of Deconfinement - CPOD 2009,\\
		 June 08 - 12 2009\\
		 Brookhaven National Laboratory, Long Island, New York, USA}

\begin{document}

%%%%%%%%%%%%%%%%%%%%%%%%%%%%%%%%%%%%%%%%%%
%%%%%%%%%%%%%%%%%%%%%%%%%%%%%%%%%%%%%%%%%%
\section{Introduction and motivation}

Although topologically nontrivial configurations of the gauge fields allow for a CP-violating term 
in the  Lagrangian of QCD \cite{Belavin:1975fg,ABJ}, experiments indicate that its numerical 
coefficient, known as $\theta$, is vanishingly small, $\theta \lesssim 10^{-10}$ \cite{exp-theta,neutron-dipole}. Nevertheless, in spite of the fact that spontaneous breaking of P and CP were proved to be 
forbidden in the true vacuum of QCD for $\theta =0$ \cite{Vafa:1984xg}, there is still room for 
metastable CP-violating states at finite temperature \cite{finitetemp}. This provides the exciting 
possibility of probing the topological structure of QCD, and leads to the idea that P- and CP-odd 
domains could be produced within the quark-gluon plasma presumably formed in high-energy heavy 
ion collisions \cite{Kharzeev:1998kz}.

The experimental signatures that were proposed to identify the presence of CP violation in certain 
domains within the plasma rely on charge separation of hadronic matter \cite{Kharzeev:1999cz}. 
This asymmetry can be enhanced to the point of becoming detectable by current experiments at 
RHIC by the presence of a very strong magnetic background, as in the case of noncentral collisions, 
generating a rich phenomenon denominated {\it chiral magnetic effect} \cite{Kharzeev:2007jp}.

The quark-gluon plasma possibly possesses regions 
with nonzero winding number, $Q_{w}\neq 0$, where sphaleron transitions are induced. 
In the particular case of noncentral collisions a strong magnetic field is generated in the orbital 
angular momentum direction, perpendicular to the reaction plane. 
In this process, the strong magnetic field ${\bf B}$ restricts the quarks (all in the lowest Landau level, aligned with {\bf B}) to move along its direction. For a topologically nontrivial domain with e.g. $Q_{w}=-1$, left-handed quarks are 
converted into right-handed ones, inducing an inversion of the direction of momentum and, 
consequently, a net current and a charge difference are created along the direction of the magnetic 
field. The system is, then, P-odd.

The magnetic fields involved, although short-lived for very high energies \cite{Kharzeev:2007jp}, 
will certainly live longer for lower values of $\sqrt{s}$, and can attain enormous intensities, above 
those considered for magnetars \cite{magnetars} and comparable only to the ones believed to be 
present during the early stages of the universe \cite{Schwarz:2003du}. The presence of such extreme 
magnetic fields coupled with strong CP violation raises several theoretical questions. The first one 
is how the QCD phase diagram is altered by the presence of a nonzero uniform magnetic field that 
plays the role of  another ``control parameter''. Next one can ask where are the possible metastable 
CP-odd states and how ``stable'' they are, i.e. how long their lifetimes are. These questions are 
directly related to possible modifications in the nature of the phase transitions of strongly interacting 
matter and to the relevant time scales for phase conversion. As we have mentioned above, there is 
already an on-going experimental investigation that seems to show indications of the chiral magnetic 
effect \cite{exp-CME}. Furthermore, pioneer lattice studies on the chiral magnetization of the 
non-Abelian vacuum also seem to be able to capture these effects \cite{lattice-CME}.

In what follows, we discuss the effects of a strong and constant magnetic background and of CP violation 
on the chiral transition at finite temperature and vanishing chemical potential. For this purpose, we 
adopt the linear sigma model coupled with two flavors of quarks as our effective field 
theory \cite{GellMann:1960np}, following the notation and conventions of Ref. \cite{Scavenius:2001bb}. 
We show that for high enough magnetic fields the chiral transition is no longer a crossover. Instead, 
it is turned into a first-order transition \cite{Fraga:2008qn}. 

To include the effects of the presence of the axial anomaly and CP violation, we add a term
that mimics the presence of nontrivial gauge field configurations, the 't Hooft
determinant \cite{'tHooft:1986nc}. The rich vacuum structure brought about by a nonzero $\theta$ 
term in the action has an influence on the chiral transition, and generates a more complex picture in 
the analysis of the phase diagram of strong interactions \cite{Boer:2008ct}. Working in a mean field 
approximation in an extended CP-odd linear sigma model, our results can be cast in terms of 
condensates of the fields \cite{Mizher:2008dj}. In this framework, three different phases occur, making the 
topography of extrema very rich, and allowing for the existence of metastable minima in certain
situations.

%%%%%%%%%%%%%%%%%%%%%%%%%%%%%%%%%%%%%%%%%%
\section{Effective theory for the chiral transition in a strong magnetic background}

Modifications in the vacuum of CP-symmetric QCD by the presence of a magnetic field
have been investigated previously within different frameworks, mainly using effective
models \cite{Klevansky:1989vi,Gusynin:1994xp,Babansky:1997zh,Klimenko:1998su,
Semenoff:1999xv,Goyal:1999ye,Hiller:2008eh,Rojas:2008sg},
especially the NJL model \cite{Klevansky:1992qe}, and chiral perturbation
theory \cite{Shushpanov:1997sf,Agasian:1999sx,Cohen:2007bt}, but also resorting to the
quark model \cite{Kabat:2002er} and certain limits of QCD \cite{Miransky:2002rp}.
Most treatments have been concerned with vacuum modifications by the magnetic field,
though medium effects were considered in a few cases, as e.g. in the study of the stability
of quark droplets under the influence of a magnetic field at finite density and zero
temperature, with nontrivial effects on the order of the chiral transition \cite{Ebert:2003yk}.
More recently, magnetic effects on the dynamical quark mass \cite{Klimenko:2008mg} and on
the thermal quark-hadron transition \cite{Agasian:2008tb}, as well as  magnetized chiral
condensates in a holographic description of chiral symmetry breaking \cite{holographic}, were
also considered. Recent applications of quark matter under strong magnetic fields to 
the physics of magnetars using the NJL model can be found in \cite{Menezes:2008qt}.

To investigate the effects of a strong magnetic background on the nature and 
dynamics of the chiral phase transition at finite temperature, $T$, and vanishing chemical 
potential, we adopt the linear sigma model coupled to two favors of quarks. 
Following the notation of Ref. \cite{Scavenius:2001bb}, we have the lagrangian
\begin{eqnarray}
{\cal L} &=&
 \overline{\psi}_f \left[i\gamma ^{\mu}\partial _{\mu} - g(\sigma +i\gamma _{5}
 \vec{\tau} \cdot \vec{\pi} )\right]\psi_f 
 %\nonumber\\
+ \frac{1}{2}(\partial _{\mu}\sigma \partial ^{\mu}\sigma + \partial _{\mu}
\vec{\pi} \partial ^{\mu}\vec{\pi} )
- V(\sigma ,\vec{\pi})\;,
\label{lagrangian}
\end{eqnarray}
where $V(\sigma ,\vec{\pi})=\frac{\lambda}{4}(\sigma^{2}+\vec{\pi}^{2} -
{\it v}^2)^2-h\sigma$ 
is the self-interaction potential for the mesons, exhibiting both spontaneous 
and explicit breaking of chiral symmetry. The $N_f=2$ massive fermion fields 
$\psi_f$ represent the up and down constituent-quark fields $\psi=(u,d)$. The 
scalar field $\sigma$ plays the role of an approximate order parameter for the 
chiral transition, being an exact order parameter for massless quarks and pions. 
The latter are represented by the pseudoscalar field $\vec{\pi}=(\pi^{0},\pi^{+},\pi^{-})$, 
and it is common to group together these meson fields into an $O(4)$ chiral field 
$\phi =(\sigma,\vec{\pi})$. 
In what follows, we implement a simple mean-field treatment with the customary simplifying 
assumptions, where quarks constitute a thermalized fluid that provides a background in which 
the long wavelength modes of the chiral condensate evolve. At $T=0$, the model reproduces 
results from chiral perturbation theory for the broken phase vacuum. 
In this phase, quark degrees of freedom are absent (excited only for $T > 0$). The $\sigma$ 
field is heavy, $M_{\sigma} \sim600$ MeV, and treated classically. On the other hand, pions are 
light, and fluctuations in $\pi^{+}$  and $\pi^{-}$ couple to the magnetic field, $B$, as will be 
discussed below, whereas fluctuations in $\pi^{0}$ give a $B$-independent 
contribution that we ignore, for simplicity. For $T > 0$, quarks are relevant (fast) degrees of 
freedom and chiral symmetry is approximately restored in the plasma for high enough $T$. 
In this case, we incorporate quark thermal fluctuations in the effective potential for $\sigma$, 
i.e. we integrate over quarks to one loop. Pions become rapidly heavy only after $T_{c}$ and 
their fluctuations can, in principle, matter since they couple to $B$. 
The parameters of the lagrangian are chosen such that the effective model reproduces 
correctly the phenomenology of QCD at low energies and in the vacuum, in the absence of 
a magnetic field.
Standard integration over the fermionic degrees of freedom to one loop,  
using a classical approximation for the chiral field, gives the effective potential 
in the $\sigma$ direction $V_{eff}= V(\phi)+V_q(\phi)$, where $V_{q}$ represents the 
thermal contribution from the quarks that acquire an effective mass $M(\sigma)=g|\sigma|$.
The net effect of the term $V_{q}$ is correcting the potential for the chiral 
field, approximately restoring chiral symmetry for a critical temperature 
$T_{c}\sim 150~$MeV \cite{Scavenius:2001bb}. 

\begin{figure}[htb]
\begin{center}
\vspace{0.7cm}
\begin{minipage}[t]{68mm}
%\framebox[79mm]{\rule[-26mm]{0mm}{52mm}}
\includegraphics[width=6.5cm]{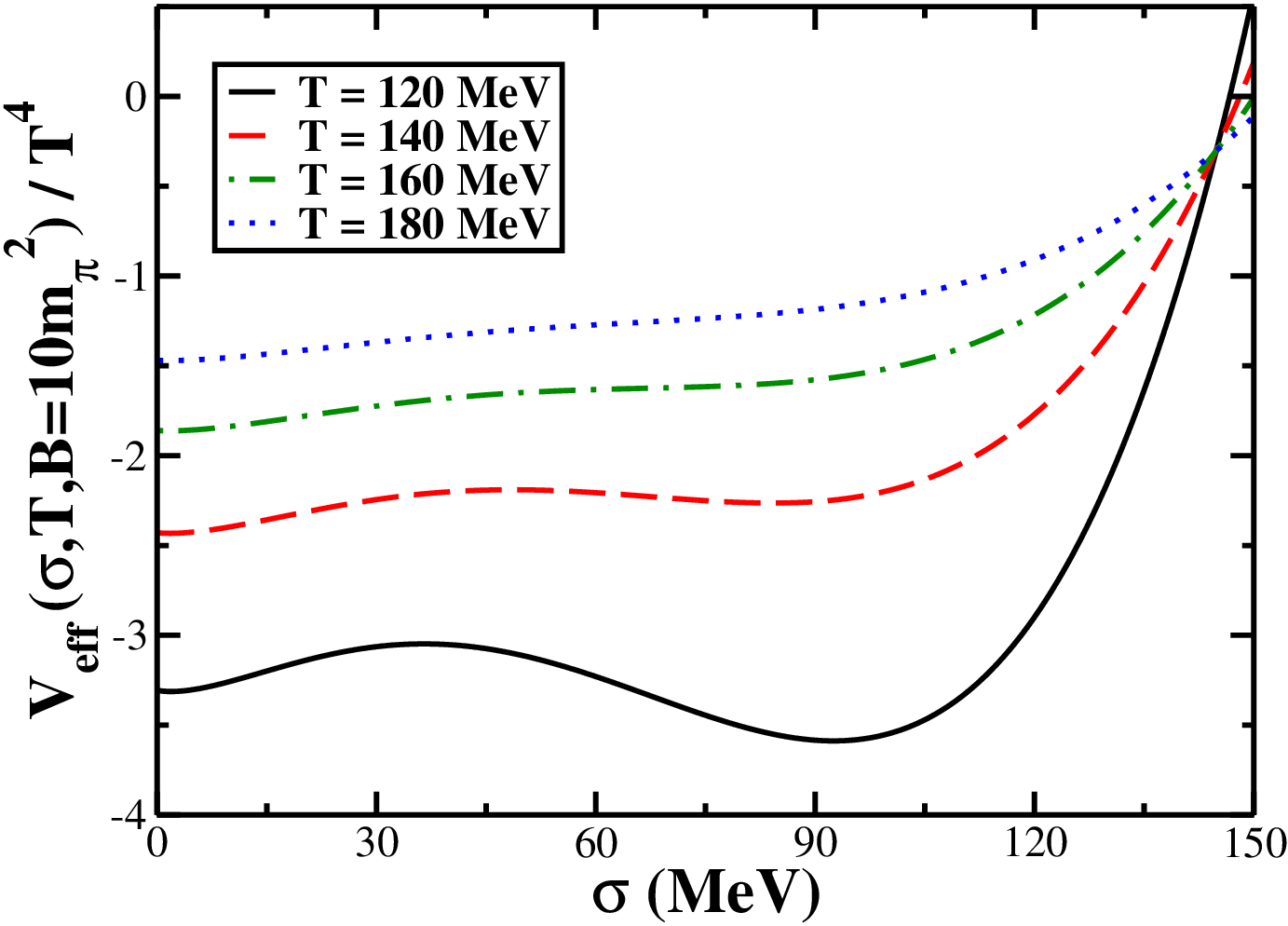}
\caption{Evidence for a first-order chiral transition in the effective potential for $eB=10 m_{\pi}^{2}$.}
\label{V}
\end{minipage}
\hspace{1cm}
\begin{minipage}[t]{68mm}
%\framebox[74mm]{\rule[-26mm]{0mm}{52mm}}
\includegraphics[width=6.7cm]{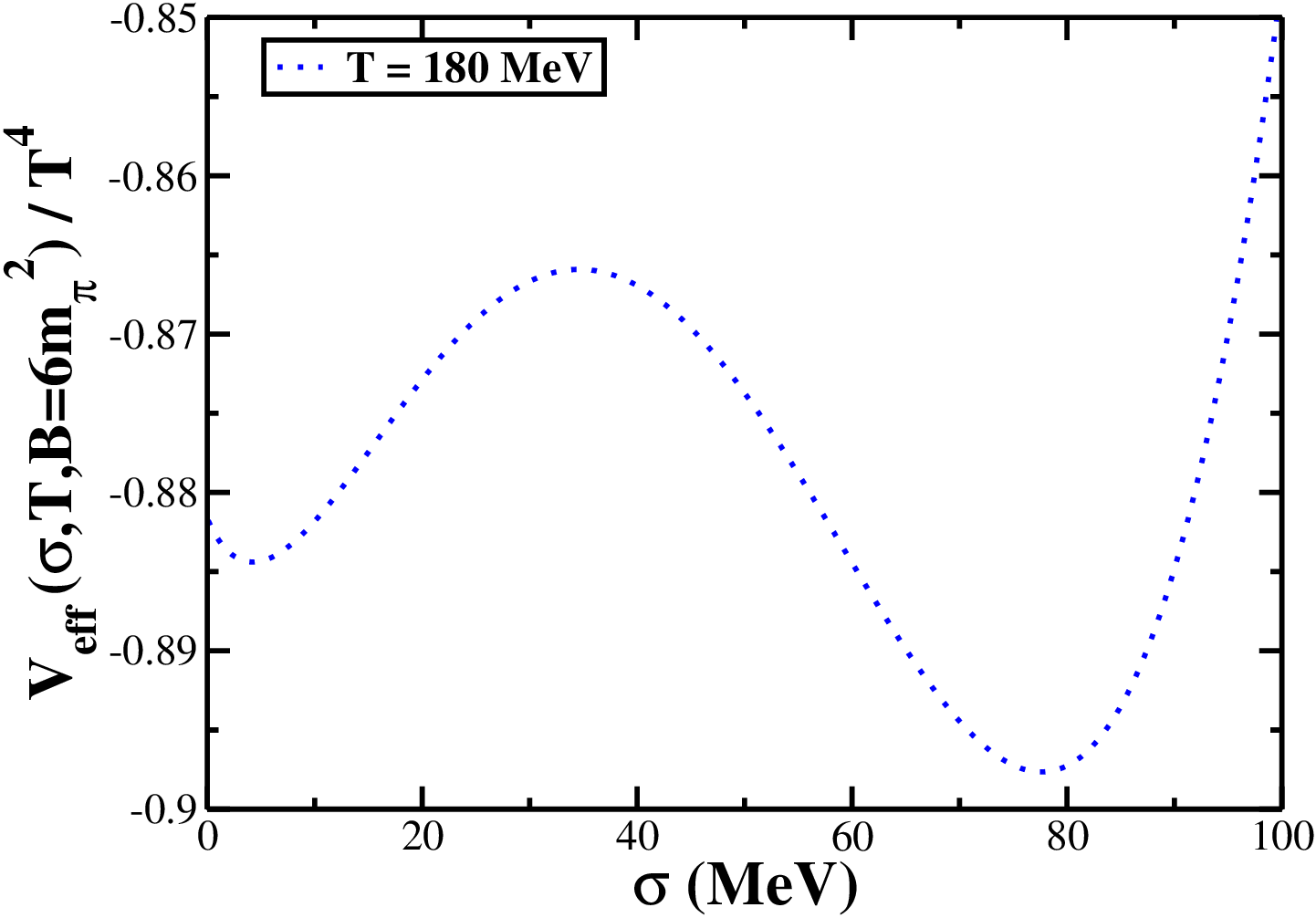}
\caption{Zoom of the barrier for $eB=6 m_{\pi}^{2}$.}
\label{Veff_B_6_zoom}
\end{minipage}
\end{center}
\end{figure}
%

%
%\begin{figure}[htb]
%\begin{center}
%\vspace{1.0cm}
%\includegraphics[width=8cm]{eff_pot_B2.eps}
%\end{center}
%\caption{Evidence for a first-order chiral transition in the effective potential for $eB=10 m_{\pi}^{2}$.}
%\label{V}
%\end{figure}
%

Assuming that the system is now in the presence of a strong magnetic background that is constant 
and homogeneous, one can compute the modified effective potential following the procedure outlined 
in Ref. \cite{Fraga:2008qn}. In what follows, we simply sketch some of the main results. 
For definiteness, let us take the direction of the magnetic field as the $z$-direction, ${\bf B}=B {\bf \hat z}$. 
The effective potential can be generalized to this case by a simple redefinition of the dispersion relations 
of the fields in the presence of ${\bf B}$, using the minimal coupling shift in the gradient and the field 
equations of motion. For this purpose, it is convenient to choose the gauge such that 
$A^{\mu}=(A^{0},{\bf A})=(0,-By,0,0)$. Decomposing the fields into their Fourier modes, one arrives 
at eigenvalue equations which have the same form as the Schr\"odinger equation for a harmonic 
oscillator potential, whose eigenmodes correspond to the well-known Landau levels. The latter 
provide the new dispersion relations
\begin{eqnarray}
p_{0n}^2=p_z^2+m^2+(2n+1)|q|B \quad , \quad
p_{0n}^2=p_z^2+m^2+(2n+1-\sigma)|q|B \, ,
\end{eqnarray}
for scalars and fermions, respectively, $n$ being an integer, $q$ the electric charge, 
and $\sigma$ the sign of the spin. Integrals over four momenta and thermal 
sum-integrals are modified accordingly, yielding sums over the Landau levels.

In our effective model, the vacuum piece of the potential will be modified by the magnetic field 
through the coupling of the field to charged pions. To one loop, and in the limit 
of high $B$, $eB >> m_{\pi}^{2}$, one obtains (ignoring contributions independent of the 
condensates) \cite{Fraga:2008qn}
\begin{equation}
V_{\pi^+}^V+V_{\pi^-}^V=-\frac{2m_\pi^2 eB}{32\pi^2}\log 2 \, .
\label{Vpion}
\end{equation}

Thermal corrections are provided by pions and quarks. However, the pion thermal contribution 
as well as part of the quark thermal contribution are exponentially suppressed for high magnetic 
fields, as has been shown in Ref. \cite{Fraga:2008qn}. The only part of the quark thermal piece 
that contributes is
\begin{equation}
V_q^T = -N_c \frac{eBT^2}{2\pi^2} \left[\int_{-\infty}^{+\infty} dx~ 
\ln\left( 1+e^{-\sqrt{x^2 +M_q^2/T^2}}\right)\right] \; ,
\label{Vquark}
\end{equation}
where $N_{c}=3$ is the number of colors. Therefore, the effective potential is corrected by the 
contributions in (\ref{Vpion}) and (\ref{Vquark}) in the presence of a strong homogeneous magnetic 
background. Therefore, the presence of the magnetic field enhances the value of the chiral condensate 
and the depth of the broken phase minimum of the modified effective potential, a result that is in line 
with those found within different approaches (see, for instance, 
Refs. \cite{Hiller:2008eh,Shushpanov:1997sf,Cohen:2007bt}).

%
%\begin{figure}[htb]
%\begin{center}
%\vspace{1.0cm}
%\includegraphics[width=8cm]{eff_pot_B6_zoom.eps}
%\end{center}
%\caption{Zoom of the barrier for $eB=6 m_{\pi}^{2}$.}
%\label{Veff_B_6_zoom}
%\end{figure}
%

Fig. 1 displays the effective potential for $eB\sim 10 m_{\pi}^{2}$ at different values of the 
temperature to illustrate the phenomenon of chiral symmetry restoration via a first-order 
transition. For RHIC top energies one expects $eB\sim 5-6 m_{\pi}^{2}$ \cite{Kharzeev:2007jp}. 
For lower values of the field, the barrier is smaller. 
In Fig. 2, we show a zoom of the effective potential for $eB\sim 6 m_{\pi}^{2}$ for a temperature 
slightly below the critical one. This figure highlights the presence of a first-order barrier in 
the effective potential. For a magnetic field of the magnitude that could 
possibly be found in non-central high-energy heavy ion collisions, one moves from a crossover 
scenario to that of a weak first-order chiral transition, with a critical temperature 
$\sim 30\%$ higher \cite{Fraga:2008qn}. Although the barrier in this case is tiny, the intensity 
of supercooling of the system is expected to be rather large due to the smallness of nucleation 
rates when compared to the expansion time scales for the heavy ion scenario. Therefore, 
even a small barrier can keep part of the system in the false vacuum until the spinodal instability 
is reached and the system is abruptly torn apart. Nevertheless, since the magnetic field falls off 
very rapidly at RHIC top energies, we expect that even for lower values of $\sqrt{s}$ only the 
early-time dynamics should be affected. 

As caveats, first we note that although non-central heavy ion collisions might show features of a 
first-order transition when contrasted to central collisions, in this comparison finite-size effects 
become important and have to be taken into account \cite{Palhares:2009tf}. Second, since the 
magnetic field varies very rapidly in time, it can induce strong electric fields that could play a 
relevant role via the Schwinger mechanism.

%%%%%%%%%%%%%%%%%%%%%%%%%%%%%%%%%%%%%%%%%%
\section{CP-odd linear sigma model}

To describe the chiral phase structure of strong interactions including CP-odd effects,
we adopt an effective model that reproduces the symmetries of QCD at low energy scales,
and has the appropriate degrees of freedom at each scale: the CP-odd linear sigma
model coupled with two flavors of quarks. The chiral mesonic sector is built including all
Lorentz invariant terms allowed by symmetry and renormalizability. Following
Refs. \cite{'tHooft:1986nc,sigma-model}, one can write
\begin{eqnarray}
\mathcal{L}_{\chi}&=& \frac{1}{2} {\rm Tr}(\partial_\mu\phi^\dagger \partial^\mu \phi)
+ \frac{a}{2} {\rm Tr}(\phi^\dagger \phi)
- \frac{\lambda_1}{4} [{\rm Tr}(\phi^\dagger \phi)]^2 
%\\ \nonumber
%&&
-\frac{\lambda_2}{4} {\rm Tr}[(\phi^\dagger \phi)^2]  \nonumber \\
&&+ \frac{c}{2}[e^{i\theta}\det(\phi) + e^{-i\theta}\det(\phi^\dagger)] 
%\\
%&&
+ {\rm Tr}[h(\phi +\phi^\dagger)] \; .
\end{eqnarray}
The potential in the Lagrangian above displays both spontaneous and explicit symmetry breaking,
the latter being implemented by the term $\sim h$. The strength of CP violation is contained in
the 't Hooft determinant term, which encodes the Levi-Civita structure of the axial anomaly and
depends on the value of the parameter $\theta$.

Expressing the chiral field $\phi$ as
\begin{equation}
\phi = \frac{1}{\sqrt{2}}(\sigma + i\eta) +
\frac{1}{\sqrt{2}} (\vec{a_0} + i \vec{\pi}) \cdot \vec{\tau} \; ,
\end{equation}
where $\vec\tau$ are the generators of $SU(2)$, the Pauli matrices, 
the potential takes the following form (substituting the parameter $h$ by $H\equiv\sqrt{2} h$):
\begin{eqnarray}
V_{\chi}&=& -\frac{a}{2} (\sigma^2 +\vec\pi^2 + \eta^2 + \vec a_0^2) 
% \\ \nonumber
%&&
-\frac{c}{2} \cos\theta ~(\sigma^2  +\vec\pi^2 - \eta^2 - \vec a_0^2) 
%\\ \nonumber
%&&
+ c ~\sin\theta ~(\sigma\eta - \vec\pi \cdot\vec a_0) \nonumber \\
&&- H\sigma 
%\\ \nonumber
%&&
+ \frac{1}{4}(\lambda_1 + \frac{\lambda_2}{2}) (\sigma^2 + \eta^2 + \vec\pi^2 + \vec a_0^2)^2 
%\\
%&&
+ \frac{2\lambda_2}{4}(\sigma\vec a_0 + \eta\vec\pi + \vec \pi \times\vec a_0)^2 \; ,
\end{eqnarray}
where the parameters $a$, $c$, $H$, $\lambda_{1}$ and $\lambda_{2}$ are fixed by vacuum
properties of the mesons \cite{Mizher:2008dj}. Quarks are coupled to the chiral fields in the same 
fashion as before.

Following a mean field analysis for the condensates $\langle\sigma \rangle$ and 
$\langle\eta \rangle$ and assuming that the remaining condensates vanish, we can compute 
the effective potential, which is a function of the condensates above and of the CP violation 
coefficient $\theta$, and fix all the free parameters \cite{Mizher:2008dj}. For $\theta =0$ the 
model, of course, reproduces the results from the usual linear sigma model. Increasing 
the value of $\theta$ from zero to $\pi$, the minima of the effective potential rotate from 
the $\sigma$ direction almost to the $\eta$ direction, the rotation being complete only 
for massless quarks.

\begin{figure}[htb]
\begin{center}
\vspace{0.5cm}
\begin{minipage}[t]{68mm}
%\framebox[79mm]{\rule[-26mm]{0mm}{52mm}}
\includegraphics[width=6.5cm]{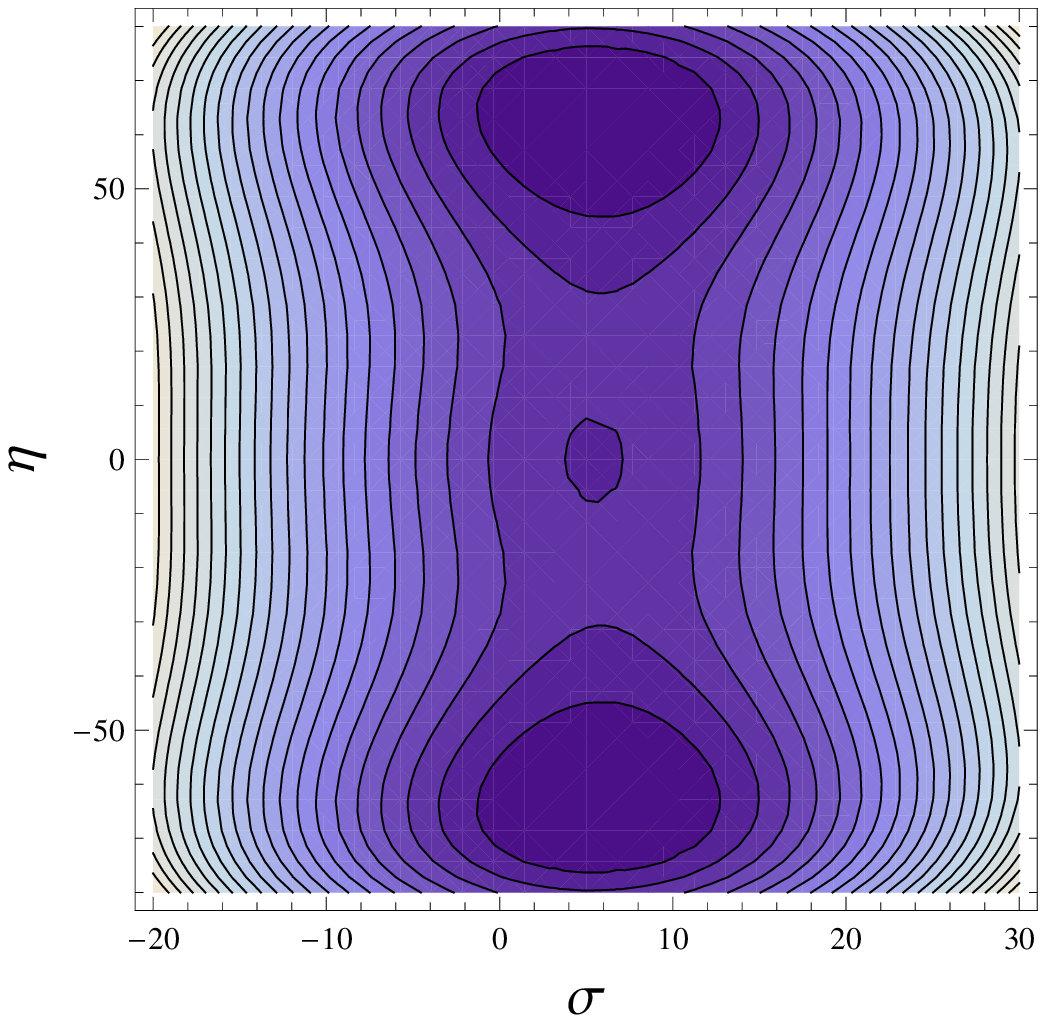}
\caption{Contour plot of the effective potential for $\theta=\pi$ and 
$T=125~$MeV. Numerical values are in MeV.}
\end{minipage}
\hspace{1cm}
\begin{minipage}[t]{68mm}
%\framebox[74mm]{\rule[-26mm]{0mm}{52mm}}
\includegraphics[width=6.4cm]{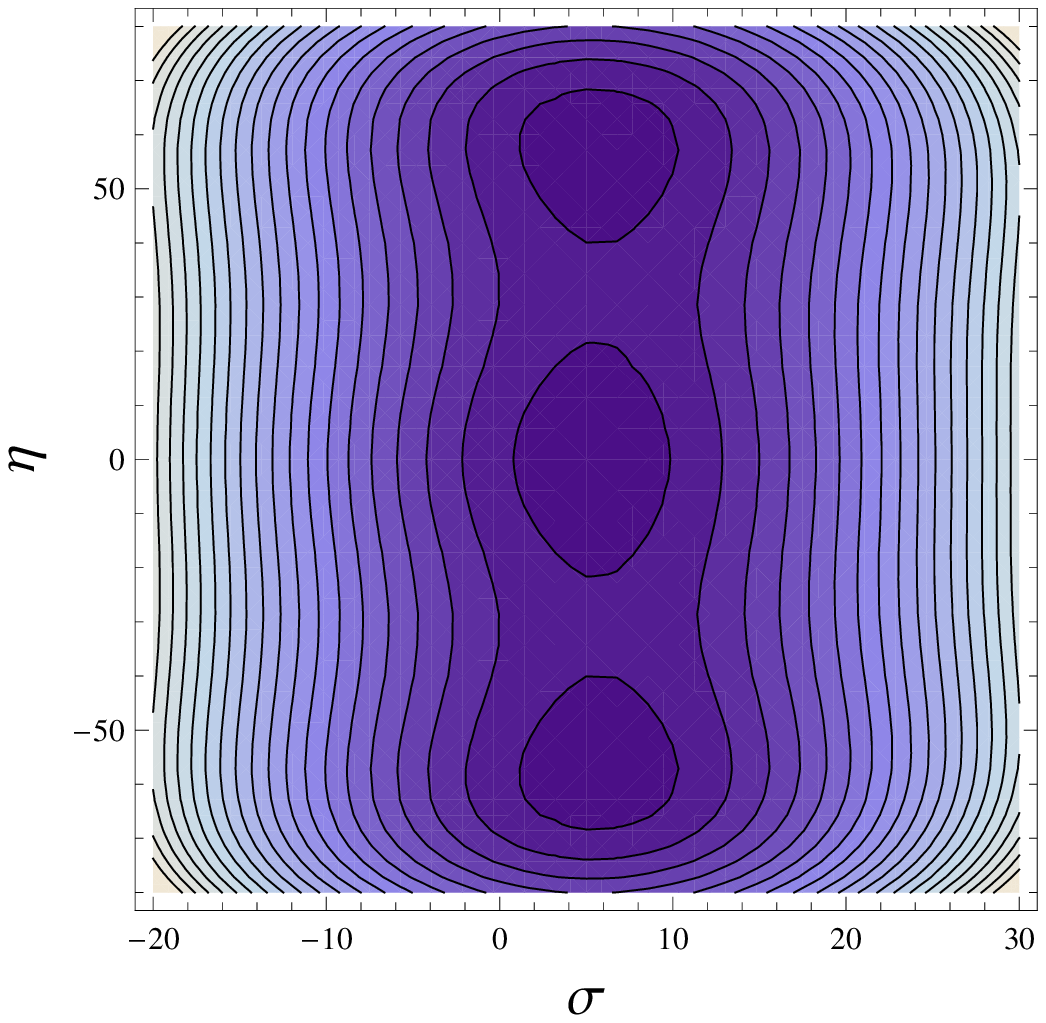}
\caption{Contour plot of the effective potential for $\theta=\pi$ and 
$T=128~$MeV. Numerical values are in MeV.}
\end{minipage}
\end{center}
\end{figure}
%

%
%\begin{figure}[!ht]
%\begin{center}
%\includegraphics[width=6cm]{CN_thetapi_T125.eps}
% \pdffig{file=CN_thetapi_T125.pdf,width=4.0cm,height=100pt}
 %\caption{Contour plot of the effective potential for $\theta=\pi$ and 
%$T=125~$MeV. Numerical values are in MeV.}
%\end{center}
%\end{figure}
%
%
%\begin{figure}[!ht]
%\begin{center}
%\vspace*{.2cm}
%\includegraphics[width=6cm]{CN_thetapi_T128.eps}
% \pdffig{file=CN_thetapi_T128.pdf,width=4.0cm,height=100pt}
 %\caption{Contour plot of the effective potential for $\theta=\pi$ and 
%$T=128~$MeV. Numerical values are in MeV.}
%\label{fig:theta=pi}
%\end{center}
%\end{figure}
%

In Figs. 3 and 4 we show contour plots of the effective potential in the case of $\theta=\pi$. 
Increasing the temperature, the minima move towards the center, indicating chiral symmetry restoration. 
However, there is a clear barrier between the global minimum and the new minimum that becomes 
the true global minimum at high temperature (at $\eta=0$). In contrast to the case with $\theta=0$, 
this signals a first-order transition, and the possibility of metastable CP-odd states. 
Moreover, since the critical temperatures for the melting of $\sigma$ and $\eta$ condensates are 
different, three different phases are allowed in systems with $\theta$ between zero and $\pi$: one in 
which both condensates are present, another where the $\eta$ condensate vanishes, and a 
phase where both condensates vanish \cite{Mizher:2008dj}. This can also be illustrated by the 
behavior of each condensate as a function of the temperature, as shown in Fig. 5.

\begin{figure}[htb]
\begin{center}
\vspace{0.75cm}
\begin{minipage}[t]{68mm}
%\framebox[79mm]{\rule[-26mm]{0mm}{52mm}}
\includegraphics[width=6.5cm]{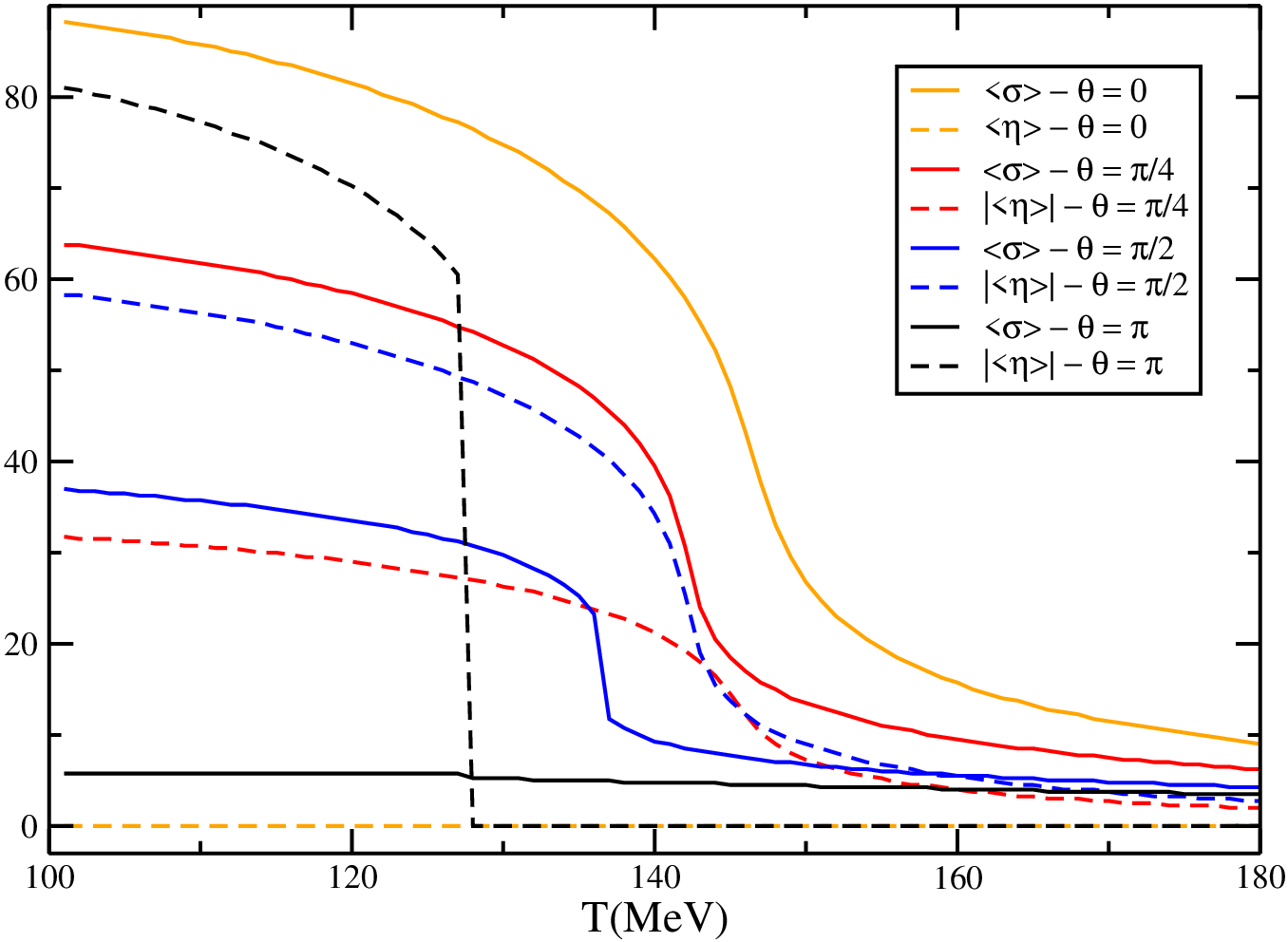}
\caption{Absolute value of the condensates (in MeV) as functions of
the temperature. Full lines denote the $\sigma$ condensate and
dotted lines the $\eta$ condensate. }
\end{minipage}
\hspace{1cm}
\begin{minipage}[t]{68mm}
%\framebox[74mm]{\rule[-26mm]{0mm}{52mm}}
\includegraphics[width=6.6cm]{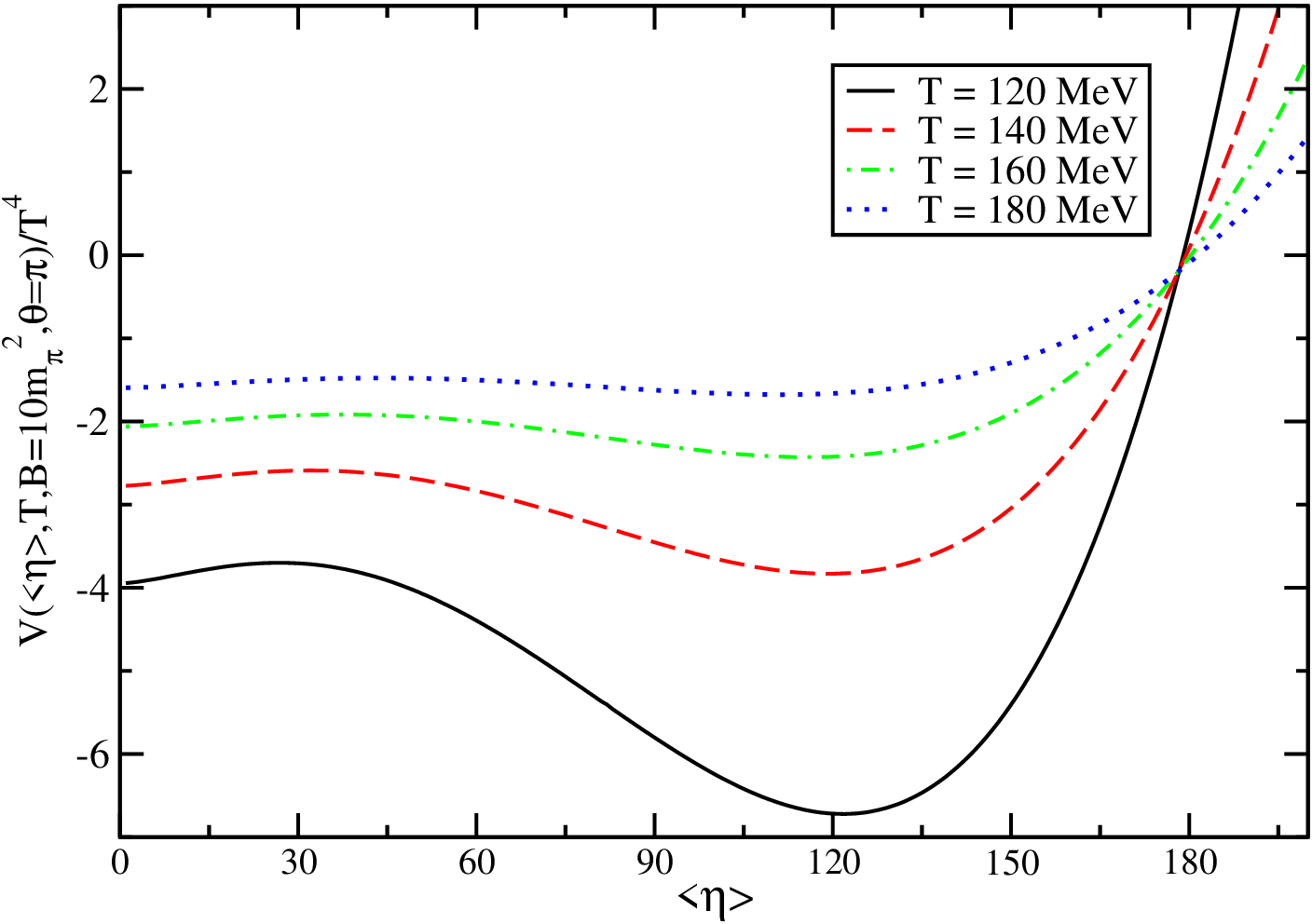}
\caption{Effective potential normalized by the temperature for $\theta=\pi$ and $B=10m_\pi^2$ 
in the direction of the $\eta$ condensate (in MeV).}
\end{minipage}
\end{center}
\end{figure}
%

%
%\begin{figure}[htbp]
%\begin{center}
%\vspace{0.3cm}
%\includegraphics[width=6cm]{sig_mod_eta_T.eps}
%\caption{Absolute value of the condensates (in MeV) as functions of
%the temperature. Full lines denote the $\sigma$ condensate and
%dotted lines the $\eta$ condensate. } 
%\label{fig:minima}
%\end{center}
%\end{figure}
%

Following the steps of the previous section, we can incorporate the effects from a strong 
magnetic background on top of the CP-odd linear sigma model \cite{Mizher:2008dj}. One 
then finds that the critical temperature is raised, as well as the barrier along the $\eta$ 
direction. Effects on the $\sigma$ field are analogous to the ones obtained in the 
CP-conserving linear sigma model. In Fig. 6, we illustrate this phenomenon for a magnetic 
field of $eB= 10 m_{\pi}^{2}$ and $\theta=\pi$.

Therefore, for nonzero $\theta$ metastable CP-violating states appear quite naturally in the CP-odd 
linear sigma model. (However, they were not found in an extension of the NJL 
model \cite{Boer:2008ct}. For a discussion, see Ref. \cite{Boomsma:2009eh}.) In fact, larger 
values of $\theta$ tend to produce a first-order chiral transition and might lead to the formation 
of domains (bubbles) in the plasma that exhibit CP violation \cite{Mizher:2008dj}. This reinforces 
the scenario of possible metastable CP-odd states in QCD that are so relevant for the chiral 
magnetic effect \cite{Kharzeev:2007jp}. This behavior is enhanced by the presence of a strong 
magnetic field, so that both effects seem to push in the same direction \cite{Fraga:2008qn,Mizher:2008dj}. 

%%%%%%%%%%%%%%%%%%%%%%%%%%%%%%%%%%%%%%%%%%
\section{Final remarks}

Strong magnetic fields can modify the nature of the chiral and the deconfining transitions, 
opening new possibilities in the study of the phase diagram of QCD, introducing a new 
control parameter, besides temperature and baryon chemical potential, in the study of the 
thermodynamics of strong interactions. These high magnetic fields are also essential in the 
context of high-energy heavy ion collisions in generating a measurable charge asymmetry to 
determine the presence of CP-odd domains created by sphaleron transitions. Several relevant 
questions that are still not fully covered can be raised. How strong can one make these magnetic fields 
in current and future experiments? How long lived could they be, and how uniform, for different 
values of $\sqrt{s}$? Only the first theoretical estimates \cite{Kharzeev:2007jp,Skokov:2009qp} and preliminary 
lattice simulations \cite{lattice-CME} were performed, providing encouraging results.

On the one hand, due to the presence of a strong magnetic background, non-central heavy ion 
collisions might show features of a first-order transition when contrasted to central collisions. On 
the other hand, finite-size effects are sizable for non-central collisions \cite{Palhares:2009tf}, so 
that an accurate centrality dependence study seems to be necessary. This demands a thin 
binning of centrality and control of finite-size effects, a very difficult (but necessary) task for 
experimentalists in the process of data analysis, especially in the scheduled Beam Energy Scan 
program at RHIC-BNL.

Finally, one still needs to perform dynamical investigations within the chiral magnetic effect scenario 
to determine the relevant time scales and verify whether effects from the CP-odd domains survive 
long enough to produce measurable signatures.

%%%%%%%%%%%%%%%%%%%%%%%%%%%%%%%%%%%%%%%%%%
\section*{Acknowledgments}
We thank D. Boer, J. Boomsma, M. Chernodub, K. Fukushima, 
M. Stephanov and A. Zhitnitsky  for fruitful discussions. E.S.F. is specially 
grateful to D. Kharzeev for valuable discussions and for his kind hospitality 
in the Nuclear Theory Group at BNL. 
This work was partially supported by CAPES, CNPq, FAPERJ and FUJB/UFRJ.

%%%%%%%%%%%%%%%%%%%%%%%%%%%%%%%%%%%%%%%%%%
%%%%%%%%%%%%%%%%%%%%%%%%%%%%%%%%%%%%%%%%%%

\end{document}